\documentclass[aps,10pt,prd,twocolumn,showpacs,preprintnumbers,nofootinbib,superscriptaddress]{revtex4-1}
\usepackage{amsmath,amssymb}
\usepackage{hyperref}
\usepackage[capitalise]{cleveref}
\usepackage{mathrsfs}
\usepackage{braket}
\usepackage{graphicx}
\usepackage{bm}
\usepackage{color}
\usepackage{ulem}

\crefname{section}{Sec.}{Secs.}

\newcommand{\tr}{\operatorname{tr}}

\begin{document}
\preprint{CHIBA-EP-237, KEK Preprint 2018-84, 2019.04.24}

\title{How to extract the dominant part of the Wilson loop average in higher representations}

\author{Ryutaro Matsudo}
\affiliation{Department of Physics, 
Faculty of Science and Engineering, 
Chiba University, Chiba 263-8522, Japan}
        \email[Electronic address: ]{afca3071@chiba-u.jp}

\author{Akihiro Shibata}
\affiliation{Computing Research Center, High Energy Accelerator Research Organization (KEK) 
SOKENDAI (The Graduate University for Advanced Studies), Tsukuba 305-0801, Japan}
        \email[Electronic address: ]{akihiro.shibata@kek.jp}

\author{Seikou Kato}
\affiliation{Oyama National College of Technology, Oyama, Tochigi 323-0806, Japan}
        \email[Electronic address: ]{skato@oyama-ct.ac.jp}

\author{Kei-Ichi Kondo}
\affiliation{Department of Physics,  
Graduate School of Science, 
Chiba University, Chiba 263-8522, Japan}
        \email[Electronic address: ]{kondok@faculty.chiba-u.jp}

\begin{abstract}
In previous works, we have proposed a new formulation of Yang-Mills theory on the lattice so that the so-called restricted field obtained from the gauge-covariant decomposition plays the dominant role in quark confinement. 
This framework improves the Abelian projection in the gauge-independent manner.
For quarks in the fundamental representation, we have demonstrated  some numerical evidences for the restricted field dominance in the string tension, which means that the string tension extracted from the restricted part of the Wilson loop reproduces the string tension extracted from the original Wilson loop.
However, it is known that the restricted field dominance is not observed for the Wilson loop in higher representations if the restricted part of the Wilson loop is extracted by adopting the Abelian projection or the field decomposition naively in the same way as in the fundamental representation.
In this paper, therefore, we focus on confinement of quarks in higher representations. 
By virtue of the non-Abelian Stokes theorem for the Wilson loop operator, we propose suitable gauge-invariant operators constructed from the restricted field to reproduce the correct behavior of the original Wilson loop averages for higher representations. 
Moreover, we perform lattice simulations to measure the static potential for quarks in higher representations using the proposed    operators. 
We find that the proposed operators well reproduce the 
 behavior of the original Wilson loop average, namely, the linear part of the static potential with the correct value of the string tension, which overcomes the problem that occurs in naively applying Abelian-projection to the Wilson loop operator for higher representations.
\end{abstract}

\maketitle
\section{Introduction}
The dual superconductor picture is one of the most promising scenarios for quark confinement \cite{dualsuper}.
According to this picture, magnetic monopoles causing the dual superconductivity are regarded as the  dominant degrees of freedom responsible for confinement. 
However, it is not so easy to verify this hypothesis. Indeed, even the definition of magnetic monopoles in the pure Yang-Mills theory is not obvious.
If magnetic charges are naively defined from electric ones by exchanging the role of magnetic field and electric one according to the electric-magnetic duality, one needs to introduce singularities to obtain non-vanishing magnetic charges, as represented by the Dirac monopole.
For such configuration, however, the energy becomes divergent.

The most frequently used prescription avoiding this issue in defining monopoles is the \textit{Abelian projection}, which is proposed by 't Hooft 
\cite{Hooft81}.
In this method, the ``diagonal component'' of the Yang-Mills gauge field is identified with the Abelian gauge field and a monopole is defined as the Dirac monopole.
The energy density of this monopole can be finite everywhere because the contribution from the singularity of a Dirac monopole can be canceled by that of the off-diagonal components of the gauge field.
In this method, however,  one needs to fix the gauge because otherwise the ``diagonal component'' is meaningless.

There is another way to define monopoles, which does not rely on the gauge fixing.
This method is called the \textit{field decomposition} which was  proposed for the $SU(2)$ Yang-Mills gauge field  by Cho \cite{Cho80}  and Duan \& Ge \cite{DG79} independently, and later readdressed by Faddeev and Niemi \cite{FN98}, and developed by Shabanov \cite{Shabanov99} and Chiba University group \cite{KMS06,KMS05,Kondo06}.
In this method, as the name suggests, the gauge field is decomposed into two parts.
A part called the \textit{restricted field} transforms under the gauge transformation just like the original gauge field, while the other part called the \textit{remaining field } transforms like an adjoint matter.
The key ingredient in this decomposition is the Lie-algebra valued field with unit length which we call the \textit{color field}.
The decomposition is constructed in such a way that the field strength of the restricted field is ``parallel'' to the color field.
Then monopoles can be defined by using the gauge-invariant part proportional to the color field in the field strength just like the Abelian field strength in the Abelian projection. 
The definition of monopoles in this method is equivalent to that in the Abelian projection.
By this construction the gauge invariance is manifestly maintained differently from the Abelian projection.
The field decomposition was extended to $SU(N)$ ($N \ge 3$) gauge field in \cite{Cho80c,FN99a,BCK02} and \cite{KSM08}. 
See e.g. \cite{KKSS15} for a review. 

While the main advantage of the field decomposition is its gauge covariance, another advantage is that, through a version of the non-Abelian Stokes theorem (NAST) invented originally by Diakonov and Petrov \cite{DP89,DP96} and extended in a unified way in  \cite{KondoIV,KT00,KT00b,Kondo08,Kondo08b,MK15,MK16}, the restricted field naturally appear in the surface-integral representation of the Wilson loop. 
By virtue of this method, we understand how monopoles contribute to the Wilson loop at least classically.

It can be numerically examined whether or not these monopoles actually reproduce the expected infrared behavior of the original Wilson loop average, even if it is impossible to do so analytically.
For quarks in the fundamental representation, indeed, such numerical simulations were already performed within the Abelian projection using the MA gauge in $SU(2)$ and $SU(3)$ Yang-Mills theories on the lattice \cite{SY90,SS94,STW02}. 
Then it was confirmed that (i) the diagonal part extracted from the original gauge field in the MA gauge reproduces the full string tension calculated from the original Wilson loop average \cite{SY90,STW02}, which is called the \textit{Abelian dominance}, 
and that (ii) the monopole part extracted from the diagonal part of the gauge field by applying the Tousaint-DeGrand procedure \cite{DT80} mostly reproduces the full string tension \cite{SS94,STW02}, which is called the \textit{monopole dominance}.

However, it should be noted that the MA gauge in the Abelian projection breaks simultaneously the local gauge symmetry and the global color symmetry. This defect should be eliminated to obtain the physical result by giving a procedure to guarantee the gauge-invariance.  For this purpose, we have developped the lattice version \cite{IKKMSS06,KSSMKI08,SKS10,KSSK11,SKKS16,KKS15} of the reformulated Yang-Mills theory written in terms of new variables obtained by the gauge-covariant field decomposition, which  enables us to perform the numerical simulations on the lattice in such a way that both the local gauge symmetry and the global color symmetry remain intact, in sharp contrast to the Abelian projection which breaks both symmetries. In this paper we adopt the gauge-covariant decomposition method to avoid these defects of the Abelian projection, although the conventional treatment equivalent to the Abelian projection  and the MA gauge can be reproduced from the gauge-covariant field decomposition method as a special case called the maximal option. 
Moreover, the MA gauge in the Abelian projection is not the only way to recover the string tension in the fundamental representation. 
By way of the non-Abelian Stokes theorem \cite{Kondo08} for the Wilson loop operator, indeed, it was found that the different type of decomposition called the minimal option is available for $SU(3)$ and $SU(N)$ for $N \ge 4$ \cite{KSM08,KSSMKI08,SKS10}.  Even for the minimal option, we have demonstrated the restricted field dominance and monopole dominance in the string tension for quarks in the fundamental representation \cite{KSSK11,SKKS16}. See \cite{KKSS15} for a review. 
Thus, our method enables to extract various degrees of freedom to be responsible for quark confinement by combining the option of gauge-covariant field decomposition and the choice of the reduction condition, which is not restricted to the Abelian projection and the MA gauge respectively. 
In this paper, indeed, we have adopted three kinds of reduction conditions to examine the contributions from magnetic monopoles of different types.

For quarks in higher representations, however, it is known that, if the Abelian projection is naively applied to the Wilson loop in higher representations, the resulting monopole contribution does not reproduce
the string tension extracted from the original Wilson loop average \cite{DFGO96}. 
This is because, in higher representations, the diagonal part of the Wilson loop does not behave in the same way as the original Wilson loop.
For example, in the adjoint representation of $SU(2)$, the diagonal part of the Wilson loop average approaches $1/3$ for a large loop, which is obviously different from the behavior of the original Wilson loop. 
In the language of the field decomposition, this means that in higher representations, the Wilson loop for the restricted field does not behave in the same way as the original Wilson loop.
Poulis  \cite{Poulis96} heuristically found the correct way to extend the Abelian projection approach for the adjoint representation of $SU(2)$.
In his approach, the diagonal part of the Wilson loop is further decomposed into the ``charged term'' and the ``neutral term'' and then the ``charged term'' is used instead of the diagonal part.


In this paper, we propose a systematic prescription to extract the ``dominant'' part of the Wilson loop average, which can be applied to the Wilson loop operator in an arbitrary representation of an arbitrary compact gauge group. 
Here the ``dominant'' part means that the string tension extracted from this part of the Wilson loop reproduces the string tension extracted from the original Wilson loop . 
In the prescription, we further extract the ``highest weight part'' from the diagonal part of the Wilson loop or the Wilson loop for the restricted field.
This prescription 
comes from the NAST.
In order to test this proposal, we calculate numerically the ``dominant'' part of the Wilson loop for the adjoint representation of $SU(2)$ group,  and adjoint and sextet representations of $SU(3)$ group. 
The results support our claim.

This paper is organized as follows.
In \cref{field_decom}, we briefly review the field decomposition of the gauge field and the NAST for the Wilson loop operator. 
In \cref{high_rep}, we propose an operator suggested from the NAST, which is expected to reproduce the dominant part of the area law fall-off of the original Wilson loop average. 
In \cref{num_result}, we perform the numerical simulations on the lattice to examine whether or not the proposed operator exhibits the expected behavior of the Wilson loop average. 
In the final section \ref{conc}, we summarize the results obtained in this paper.
In \cref{sec:deriv_WV,sec:deriv_tildeW} we give derivation of some equations given in \cref{high_rep}

\section{Field decomposition method and the non-Abelian Stokes theorem} \label{field_decom}

In this section, we give a brief review of the field decomposition, the non-Abelian Stokes theorem (NAST) for the Wilson loop operator and the reduction conditions.
First, we introduce the field decomposition in a continuum  theory and then in a lattice  theory.
Here we work in the $SU(N)$ Yang-Mills theory, but the field decomposition can be applied to an arbitrary compact group \cite{MK16}.
{Next we introduce the Diakonov-Petrov version of the non-Abelian Stokes theorem \cite{DP89} for the Wilson loop operator, which is used to see
the relationship between the field decomposition and the Wilson loop operator. }
Finally, we explain {the relationship between the field decomposition and} the reduction condition which determines the color fields as a functional of the gauge field.
For a more detailed review, see e.g., \cite{KKSS15}.

\subsection{Field decomposition}

\subsubsection{Continuum case}

In the field decomposition method, we decompose the gauge field $\mathscr A_\mu(x)$ into two parts as
\begin{align}
  \mathscr A_\mu(x) = \mathscr V_\mu(x) + \mathscr X_\mu(x).
\end{align}
Here the restricted field $\mathscr V_\mu(x)$ is required to transform  just like the gauge field $\mathscr A_\mu$ under the gauge transformation as
\begin{align}
  \mathscr V_\mu(x) \rightarrow {g(x)}\mathscr V_\mu(x) {g^\dag(x)} + i{g_{\mathrm{YM}}^{-1}} {g(x)}\partial_\mu {g^\dag(x)},
\end{align}
where $g(x)\in SU(N)$ and $g_{\mathrm{YM}}$ is the Yang-Mills coupling.
Hence the remaining field $\mathscr X_\mu(x)$ must transform like an adjoint matter field as
\begin{align}
  \mathscr X_\mu(x) \rightarrow {g(x)}\mathscr X_\mu(x) {g^\dag(x)}.
\end{align}
We wish to regard the restricted field $\mathscr V_\mu$ as the dominant part of the gauge field $\mathscr A_\mu$ in the IR region.
In this paper, we focus on the version of maximal option.

In order to determine the decomposition for the gauge group $SU(N)$, we introduce a set of  \textit{color fields} $\bm n^{(k)}(x)$ ($k=1,\ldots,N-1$)  which are expressed  using a common $SU(N)$-valued field $\Theta(x)$ as 
\begin{align}
  \bm n^{(k)}(x) := \Theta(x) H_k \Theta^\dag(x), \label{color_lat}
\end{align}
where $H_k$ is a Cartan generator.
Notice that the color fields are not independent.
The transformation property of the color fields under a gauge transformation is given by
\begin{align}
  \bm n^{(k)}(x) \rightarrow {g(x)}\bm n^{(k)}(x) {g^\dag(x)}.
\end{align}
The color fields are determined as functionals of $\mathscr A_\mu$ by imposing a condition which we call the \textit{reduction condition} as  explicitly given shortly.

The decomposition is constructed such that the field strength of the restricted field, $\mathscr F_{\mu\nu}[\mathscr V]:= \partial_\mu\mathscr V_\nu -\partial_\nu\mathscr V_\mu -ig[\mathscr V_\mu,\mathscr V_\nu]$, is expressed by a linear combination of the color fields.
This condition can be simply written as
\begin{align}
  \mathscr D_\mu[\mathscr V]\bm n^{(k)} = 0 \qquad (k=1,\ldots,N-1) ,
\label{cond1}
\end{align}
where $\mathscr D_\mu[\mathscr V]:= \partial_\mu -ig_{\mathrm{YM}} [\mathscr V_\mu,\bullet]$ is the covariant derivative with the restricted field $\mathscr V_\mu$. 
This condition is manifestly gauge covariant.
This determines the component of the restricted field orthogonal to the Lie subalgebra spanned by the color fields, but does not determine the component parallel to it. 
Therefore we need to impose another condition.
We wish to identify the restricted field with the dominant part of the original gauge field, and thus it should be as close as possible to the original gauge field in the IR region.
For this reason we impose the condition that the component of the restricted field parallel to the color fields is the same as that of the gauge field as
\begin{align}
  &\operatorname{tr}(\bm n^{(k)}\mathscr V_\mu) = \tr(\bm n^{(k)}\mathscr A_\mu) \qquad (k=1,\ldots,N-1) .
\label{cond2}
\end{align}
The two conditions \cref{cond1,cond2} uniquely determine the decomposition
as
\begin{align}
  \mathscr V_\mu &= \sum_{k=1}^{N-1}2\operatorname{tr}(\bm n^{(k)}\mathscr A_\mu)\bm n^{(k)}  -i{g_{\mathrm{YM}}^{-1}}\sum_{k=1}^{N-1}[\bm n^{(k)},\partial_\mu \bm n^{(k)}], \notag\\
  \mathscr X_\mu &= i{g_{\mathrm{YM}}^{-1}}\sum_{k=1}^{N-1}[\bm n^{(k)}, \mathscr D_\mu[\mathscr A]\bm n^{(k)}]. \label{cont_decom}
\end{align}
In fact, the resulting decomposed fields satisfy the required transformation properties.
As the field strength $\mathscr F_{\mu\nu}[\mathscr V]$ {can be written as the linear combination of the color fields}, we can define Abelian-like gauge-invariant field strength as
\begin{align}
  F^{(k)}_{\mu\nu} := 2\operatorname{tr}(\bm n^{(k)} \mathscr F_{\mu\nu}[\mathscr V]), \label{def_F}
\end{align}
where the normalization of the Cartan generators is given as $\operatorname{tr}(H_kH_l) = \delta_{kl}/2$.
Then monopoles are defined in the same manner as the Dirac monopoles for this field strength $F^{(k)}_{\mu\nu}$.
The resulting monopoles are gauge invariant by construction.

{The color fields $\bm n^{(k)}$ are obtained by imposing a reduction condition as we said before.
If a reduction condition is given by minimizing a functional}
\begin{align}
  &R_{\mathrm{MA}}[\mathscr A,\{\bm n^{(k)}\}]\notag\\
  &= \int d^D x \sum_{k=1}^{N-1}\operatorname{tr}(\mathscr D_\mu[\mathscr A]\bm n^{(k)}(x)\mathscr D_\mu[\mathscr A]\bm n^{(k)}(x)), 
\label{cont_red_con}
\end{align}
the definition of monopoles is equivalent to that for the Abelian projection in the MA gauge.

\subsubsection{Lattice case} 

In the lattice version of the field decomposition \cite{IKKMSS06,KSSMKI08,SKS10}, 
a link variable $U_{x,\mu}$ is decomposed into two variables as
\begin{align}
  U_{x,\mu} = X_{x,\mu}V_{x,\mu}, \qquad X_{x,\mu},V_{x,\mu}\in SU(N),
\end{align}
where $V_{x,\mu}$ gauge-transforms just like a link variable as
\begin{align}
  V_{x,\mu} \rightarrow {g_x} V_{x,\mu} {g_{x+\mu}^\dag}, \qquad {g_x} \in SU(N), \label{trans_V}
\end{align}
and $X_{x,\mu}$ transforms like an adjoint matter as
\begin{align}
  X_{x,\mu} \rightarrow {g_x} X_{x,\mu} {g_x^\dag}.
\end{align}
The decomposition is determined by using the color fields $\bm n_x^{(k)} = \Theta_x H_k \Theta_x^\dag$ ($k=1,\ldots,N-1$) in the similar way to the continuum case.
The first condition which determines the decomposition is given by replacing the covariant derivative $\mathscr D_\mu[\mathscr V]$ in \cref{cond1} with the covariant lattice derivative $D_\mu[V]$ as 
\begin{align}
  D_\mu[V]\bm n_x^{(k)} := \varepsilon^{-1}(V_{x,\mu}\bm n_{x+\mu}^{(k)} - \bm n_{x}^{(k)}V_{x,\mu}) = 0, \label{cond1lat}
\end{align}
where $\varepsilon$ is the lattice spacing.
This condition does not determine $V_{x,\mu}$ completely because this equality is maintained if we multiply $V_{x,\mu}$ from the left by $g_x\in SU(N)$ which satisfies $[\bm n_x^{(k)}, g_x]=0$ for any $k$.
To reproduce the continuum version of the decomposition \cref{cont_decom} in the naive continuum limit, the decomposition is chosen as \cite{SKS10}
\begin{align}
  V_{x,\mu} &= \hat K_{x,\mu} U_{x,\mu}(\operatorname{det}(\hat K_{x,\mu}))^{-1/N}, \notag\\
   X_{x,\mu} &= \hat K^\dag_{x,\mu}(\operatorname{det}(\hat K_{x,\mu}))^{1/N}, \notag\\
  \hat K_{x,\mu} &:= \left(\sqrt{K_{x,\mu}K^\dag_{x,\mu}}\right)^{-1}K_{x,\mu}, \notag\\
  K_{x,\mu} &:= \bm 1 + 2N\sum_{k=1}^{N-1}\bm n_x^{(k)}U_{x,\mu}\bm n^{(k)}_{x+\mu}U^\dag_{x,\mu} . \label{lat_decom}
\end{align}

The color fields are determined by minimizing a reduction functional as in the continuum case.
The lattice version of \cref{cont_red_con} is given by replacing the covariant derivative with the covariant lattice derivative as
\begin{align}
  R_{\mathrm{MA}}[U,\{\bm n^{(k)}\}] := \sum_{x,\mu}\sum_{k=1}^{N-1} \tr[(D_\mu[U]\bm n^{(k)}_x)^\dag D_\mu[U]\bm n^{(k)}_x]. \label{MA}
\end{align}

%
%
\subsection{Non-Abelian Stokes theorem}

The Wilson loop operator in a representation $R$ is defined by
\begin{align}
W_R[\mathscr V;C] &:= \frac1{D_R}\tr_R \mathcal P\exp\left( i{g_{\mathrm{YM}}}\oint_C \mathscr A  \right),
  \end{align}
  where $D_R$ is the dimension of $R$, $\tr_R$ denotes the trace in $R$ and $\mathcal P$ denotes the path ordering.
We can relate the decomposed field variables to a Wilson loop operator through a version of the NAST which was proposed by Diakonov and Petrov \cite{DP89}.
In this version of the NAST, a Wilson loop operator in a representation $R$ is rewritten into the surface integral form by introducing a functional integral on the surface $S$ surrounded by the loop $C$ as
\begin{align}
  &W_R[\mathscr A;C] = \int D\Omega \exp\left( i{g_{\mathrm{YM}}} \int_{S:\partial S=C} \sum_{k=1}^{N-1}\Lambda_k F^{(k)}\right), \notag\\
  &F^{(k)} := \frac12F^{(k)}_{\mu\nu} dx^\mu\wedge dx^\nu,\qquad D\Omega := \prod_{x\in S}d\Omega(x) \label{NAST}
\end{align}
where $D\Omega$ is the product of the Haar measure $d\Omega(x)$ over the surface $S$ with the loop $C$ as the boundary, $\Lambda_k$ is the $k$-th component of the highest weight of the representation $R$, the color fields are defined by $\bm n^{(k)} = \Omega H_k \Omega^\dag$ and $F^{(k)}_{\mu\nu}$ is the Abelian-like field strength defined by \cref{def_F}.
Thus we can relate the restricted field to the Wilson loop operator in the manifestly gauge-invariant way.

The simplified version of the derivation is as follows.
See e.g., \cite{KT00,Kondo08,KKSS15} for more detailed derivation of \cref{NAST} along the following line.
First, we divide the loop into small pieces and represent the Wilson loop operator as the product of the parallel transporter for each piece.
Next we insert between parallel transporters the completeness relation
\begin{align}
  \bm 1 = \int d\Omega\,\Omega\ket{\Lambda}\bra{\Lambda}\Omega^\dag,
\end{align}
where $d\Omega$ is the Haar measure and $\ket{\Lambda}$ is the highest weight state of the representation $R$, and rewrite the trace by using the equality
\begin{align}
  \operatorname{tr}\mathscr O = \int d\Omega \bra{\Lambda}\Omega^\dag \mathscr O \Omega \ket{\Lambda}.
\end{align}
Then, by taking the limit where the length of each piece of the loop goes to zero, we obtain
\begin{align}
  &W_R[\mathscr A;C] = \int \prod_{x\in C}d\Omega(x) \exp\left( i{g_{\mathrm{YM}}}\oint_C  \bra\Lambda \mathscr A^\Omega\ket \Lambda \right), \notag\\
  &\mathscr A^\Omega(x) := \Omega^\dag(x)\mathscr A(x) \Omega(x) +i{g_{\mathrm{YM}}^{-1}} \Omega^\dag(x)d\Omega(x). \label{nast2}
\end{align}
In this expression, the path ordering disappears and therefore we can use the usual Stokes theorem as
\begin{align}
 &W_R[\mathscr A;C] = \int \prod_{x\in S}d\Omega(x) \exp\left( i{g_{\mathrm{YM}}}\int_{S:\partial S=C} F^{\Omega}\right), \notag\\
  &F^{\Omega}(x) := d\bra{\Lambda}\mathscr A^{\Omega}\ket{\Lambda}. \label{nast}
\end{align}
We can show that $F^{\Omega}_{\mu\nu}$ is written as the linear combination of the Abelian-like field strengths \cref{def_F} as \cite{MK15}
\begin{align}
  F^{\Omega}_{\mu\nu} = \sum_{k=1}^{N-1}\Lambda_k F^{(k)}_{\mu\nu},\qquad \bm n^{(k)}(x):= \Omega(x)H_k \Omega^\dag(x),
\end{align}
where the color fields $\bm n^{(k)}$ is defined by using the integration variable $\Omega(x)$ instead of $\Theta(x)$.

Clearly, the NAST can be applied not only to the fundamental representation but also to any representation, suggesting the correct way for extracting the dominant part of the Wilson loop in higher representations as we explain in the next section.

\subsection{The relationship between the NAST and the reduction condition}
Here we consider the relation between the reduction condition and the NAST.
In the NAST \cref{NAST}, we observe that the field strength $F_{\mu\nu}^{(k)}$ is defined in terms of the integration variable $\Omega(x)$.
At this stage, $\Omega(x)$ is distinct from $\Theta(x)$ {used to define the field decomposition.} Therefore, there is no clear relationship between the Wilson loop operator and the field decomposition defined by using the color field $\bm n^{(k)}(x)$ constructed from $\Theta(x)$.
Instead of performing the integration over the measure $D\Omega$, the color fields defined using $\Omega(x)$ in \cref{NAST} are replaced by the color fields defined using $\Theta(x)$ {determined by solving the reduction condition}. 

The reduction condition is not determined uniquely.
To see the dependence on the reduction condition,
   in the present study for the $SU(3)$ Yang-Mills theory, 
   we performed numerical simulations under the two additional reductions conditions which are defined by minimizing the functionals
   \begin{align}
   R_{n3}[ U,\{\bm n^{(k)}\}] &= \sum_{x,\mu}\operatorname{tr}[(D_\mu[ U]\bm n^3_x)^\dag  D_\mu[ U]\bm n^3_x], 
   \label{n3}
   \\
     R_{n8}[ U,\{\bm n^{(k)}\}] &= \sum_{x,\mu} \operatorname{tr}[ (D_\mu[ U]\bm n^8_x)^\dag  D_\mu[ U]\bm n^8_x], 
   \label{n8}
   \end{align}
   where $\bm n^3_x := \Theta_x T^3\Theta^\dag_x$ and $\bm n^8_x := \Theta_x T^8\Theta^\dag_x$.
   Note that the reduction functional \cref{n8} does not determine $\bm n^3_x$ and therefore does not determine the decomposition \cref{lat_decom} completely.
   However, as we explain in the next section, a specific part \cref{n8_determ} of the Wilson loop for the restricted field is determined.


\begin{figure*}[t]
  \centering
  \includegraphics{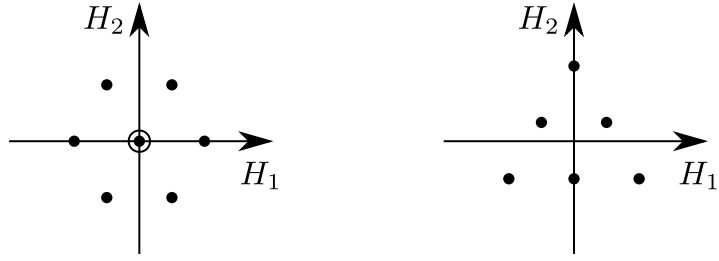}
  \caption{The weight diagram of (a) the adjoint representation $[1,1]$ and (b) the sextet representation $[0,2]$ of $SU(3)$. A single dot represents a weight $\bm \mu$ with multiplicity one, $d_{\mu}=1$, and a circled dot represents a weight $\bm \mu$ with multiplicity two, $d_{\mu}=2$.}
  \label{fig:weight}
\end{figure*}

\section{Wilson loops in higher representations} \label{high_rep}

In the preceding numerical simulations \cite{SS94,STW02} by using the Abelian projection and \cite{IKKMSS06,KSSK11,SKKS16,KKS15,CC} by using the field decomposition, it was shown that the area law of the average of a Wilson loop in the fundamental representation is reproduced by the monopole contribution.
However, this might be an accidental agreement restricted to the fundamental representation.
Therefore, we should check the other quantities.
The Wilson loops in higher representations are appropriate for this purpose because they have clear physical meaning.
However, it is known that if we apply the Abelian projection naively to higher representations, the monopole contributions in the Abelian part do not reproduce the correct behavior \cite{DFGO96}.
For example, in the adjoint representation of $SU(2)$,  the Abelian Wilson loop average approaches $1/3$ as the loop size increases according to the numerical simulation \cite{Poulis96}.
In this case, we cannot extract the static potential $V(R)$  from the exponential fall-off  behavior $e^{-V({L})T}$ of the Wilson loop average defined for the rectangular loop with length $T$ and width $L$, since $e^{-V({L})T} \to 0$ as $T \to \infty$.
In the spin-$3/2$ representation, the string tension extracted from the Abelian Wilson loop has the same value as that for the fundamental representation \cite{DFGO96}, which is different from the correct behavior.
Thus we need to find a more appropriate way to extract the monopole contributions in the Abelian part.

As we mentioned before, the NAST suggests how we extract the ``dominant part'' of the Wilson loop average, which means that by using an appropriate operator $\widetilde W_R[\mathscr V;C]$ suggested by the NAST, we can reproduce the full string tension extracted using the original Wilson loop $W_R[\mathscr A;C]$.
In the language of the field decomposition, the diagonal part of the Wilson loop is equivalent to the ``restricted Wilson loop'' $W_R[\mathscr V;C]$, the Wilson loop for the restricted field $\mathscr V$.
Therefore, the average of $W_R[\mathscr V;C]$ does not reproduce the string tension extracted from the original Wilson loop $W_R[\mathscr A;C]$.
On the other hand, the NAST \cref{NAST} suggests the distinct operator $\widetilde W_R[\mathscr V;C]$ as the dominant part of the Wilson loop in higher representations.

We now give the explicit expressions for the operators suggested by the NAST, $\widetilde W_R[\mathscr V;C]$,
and the restricted Wilson loop operator  $W_R[\mathscr V;C]$ to see the difference between the two operators. 
The restricted Wilson loop operator  $W_R[\mathscr V;C]$ is rewritten as
\begin{align}
  W_R[\mathscr V;C] &:= \frac1{D_R}\tr_R \mathcal P\exp\left( i{g_{\mathrm{YM}}}\oint_C \mathscr V  \right) \notag\\
  &= \frac1{D_R}\sum_{\bm\mu\in \Delta_R} d_\mu \exp\left(i{g_{\mathrm{YM}}}\oint_C \bra\mu\mathscr A^{\Theta}\ket\mu \right), \label{WV}
\end{align}
where $D_R$ is the dimension of the representation $R$, $\Delta_R$ is the set of all weights of $R$, $d_\mu$ is the multiplicity of a weight $\bm \mu$ and $\ket\mu$ is a normalized state corresponding to $\bm \mu$.
Note that this operator \cref{WV} is gauge invariant just as the original Wilson loop.
The derivation of \cref{WV} is given in \cref{sec:deriv_WV}.

For example, in the adjoint representation of $SU(2)$, the Wilson loop for the restricted field is written as
\begin{align}
  &W_{J=1}[\mathscr V;C]
  = \frac13\left( e^{i\phi} + e^{-i\phi} + 1 \right), \notag\\
  &\phi := {g_{\mathrm{YM}}}\oint 2\tr(\mathscr A^\Theta T^3).
\end{align}
In \cite{Poulis96}, it was confirmed that the average of this operator approaches $1/3$ as the loop size increases.
This behavior is clearly different from the original Wilson loop.

In the adjoint representation $[1,1]$ and the sextet representation $[0,2]$ of $SU(3)$, the weight diagram is given in \cref{fig:weight} (a) and (b) respectively. Then the Wilson loop for the restricted field is written as
\begin{align}
  &W_{[1,1]}[\mathscr V;C]
  = \frac18\left( e^{i\frac{\phi_3 + \sqrt3\phi_8}2} + e^{-i\frac{\phi_3 + \sqrt3\phi_8}2}
  + e^{i\frac{-\phi_3 + \sqrt3\phi_8}2}\right. \notag\\
  &\hphantom{W_{[1,1]}[\mathscr V;C]}\quad\left.+ e^{-i\frac{-\phi_3 + \sqrt3\phi_8}2}
  + e^{i\phi_3} + e^{-i\phi_3} + 2\right), \notag\\
  &W_{[0,2]}[\mathscr V;C]
  = \frac16\left( e^{i\frac2{\sqrt3}\phi_8} + e^{i\frac{3\phi_3+\sqrt3\phi_8}3}+e^{i\frac{-3\phi_3+\sqrt3\phi_8}3}\right. \notag\\
  &\hphantom{W_{[1,1]}[\mathscr V;C]}\quad \left.+ e^{i\frac{3\phi_3+\sqrt3\phi_8}6} + e^{i\frac{-3\phi_3+\sqrt3\phi_8}6} + e^{-i\frac1{\sqrt3}\phi_8}\right), \notag\\
  &\phi_3 := {g_{\mathrm{YM}}}\oint_C2\tr(\mathscr A^{\Theta}T^3), \
  \phi_8 := {g_{\mathrm{YM}}}\oint_C2\tr(\mathscr A^{\Theta}T^8).
\end{align}

On the other hand, the operator $\widetilde W_R[\mathscr V;C]$ suggested by the NAST is the integrand of the NAST using the color fields satisfying the reduction condition, i.e., the integrand of \cref{nast2} with $\Omega(x) = \Theta(x)$.
We include the contribution of the weights which are equivalent to the highest weight under the action of the Weyl group.
Let the set of such weights be $\Delta_R^h$.
Thus we propose the operator
\begin{align}
 \widetilde W_R[\mathscr V;C] &= \frac1{D_R^h}\sum_{\bm\Lambda\in\Delta^h_R} \exp\left( i{g_{\mathrm{YM}}}\oint_C \bra\Lambda \mathscr A^{\Theta}\ket\Lambda \right), \label{tildeW}
\end{align}
where 
$D_R^h$ is the number of elements in $\Delta^h_R$.
We call this operator as \textit{the highest weight part} of the Abelian Wilson loop.
Note that this operator \cref{tildeW} is gauge invariant because $\Theta(x)$ transforms as $\Theta(x)\rightarrow {g(x)}\Theta(x)$ under the gauge transformation.
In the fundamental representation, the highest weight part of the Abelian Wilson loop, \cref{tildeW}, is the same as the Abelian Wilson loop because all weights of the fundamental representation is equivalent to the highest weight under the action of the Weyl group.

For example, in the adjoint representation of $SU(2)$ the proposed operator is written as
\begin{align}
  \widetilde W_{J=1}[\mathscr V;C]
  = \frac12\left(e^{i\phi} + e^{-i\phi} \right).
\end{align}
In \cite{Poulis96}, Poulis heuristically found that this operator reproduces the full adjoint string tension without giving the theoretical justification.
In the adjoint representation $[1,1]$ and the sextet representation $[0,2]$ of $SU(3)$ it can be written as
\begin{align}
  \widetilde W_{[1,1]}[\mathscr V;C]
  &= \frac16\left( e^{i\frac{\phi_3 + \sqrt3\phi_8}2} + e^{-i\frac{\phi_3 + \sqrt3\phi_8}2}
  + e^{i\frac{-\phi_3 + \sqrt3\phi_8}2}\right. \notag\\
  &\quad\left.+ e^{-i\frac{-\phi_3 + \sqrt3\phi_8}2}
  + e^{i\phi_3} + e^{-i\phi_3} \right), \notag\\
  \widetilde W_{[0,2]}[\mathscr V;C]
  &= \frac13\left( e^{i\frac{2}{\sqrt3}\phi_8} + e^{i\frac{3\phi_3+\sqrt3\phi_8}3} + e^{i\frac{-3\phi_3+\sqrt3\phi_8}3}\right).
\end{align}

For $SU(2)$, the proposed operator \cref{tildeW} for the spin-$J$ representation can be written as
\begin{align}
  \widetilde W_J[V;C] = \frac12 \tr \left( (V_C)^{2J} \right),
\label{su2}
\end{align}
by using the untraced restricted Wilson loop $V_C$ in the fundamental representation defined as
\begin{align}
 V_C:= \prod_{\braket{x,\mu}\in C}V_{x,\mu}.
\end{align}
For $SU(3)$,  the proposed operator for the representation with the Dynkin index $[m,n]$ can be written as 
\begin{align}
  &\widetilde W_{[m,n]}[V;C] \notag\\
  &= \frac16\left(\tr\left((V_C)^{m}\right) \tr\left((V_C^{\dag})^{n}\right) - \tr\left((V_C)^{m}(V_C^{\dag})^{n}\right)\right), \label{su3}
\end{align}
where $(V_C)^0 = \bm 1$.
The derivation of \cref{su2,su3} is given in \cref{sec:deriv_tildeW}.
Note that \cref{su2,su3} are gauge invariant because of the gauge-transformation property of $V_{x,\mu}$, \cref{trans_V}.
Indeed \cref{su2} for $J=1/2$ in $SU(2)$ and \cref{su3} for $[m,n]=[1,0]$ in $SU(3)$ are the same as the ordinary Abelian Wilson loop in the fundamental representation.

Finally, we consider what part of the Abelian Wilson loop is determined by the reduction condition \cref{n8}.
The color field $\bm n_x^8$ does not change under a transformation $\Theta_x \rightarrow \Theta_xg_x$, $g_x\in U(2)$, where $U(2)$ is generated by $T^1,T^2,T^3,T^8$.
Under this transformation $\phi_8$ does not change but $\phi_3$ changes.
Thus a part of the Abelian Wilson loop which is determined by \cref{n8} is written as
\begin{align}
  e^{i\frac n{\sqrt3}\phi_8},\qquad n\in\bm Z. \label{n8_determ}
\end{align}
This part is contained in the highest weight part of the Abelian Wilson loop {only} for representations $[m,0]$ and $[0,n]$.
Therefore, in the numerical simulation, we have not calculated the highest weight part of the Abelian Wilson loop in the adjoint representation $[1,1]$ for the reduction condition \cref{n8}.

\section{Numerical result} \label{num_result}

\begin{figure}[t]
\centering
\includegraphics[width=0.95\hsize]{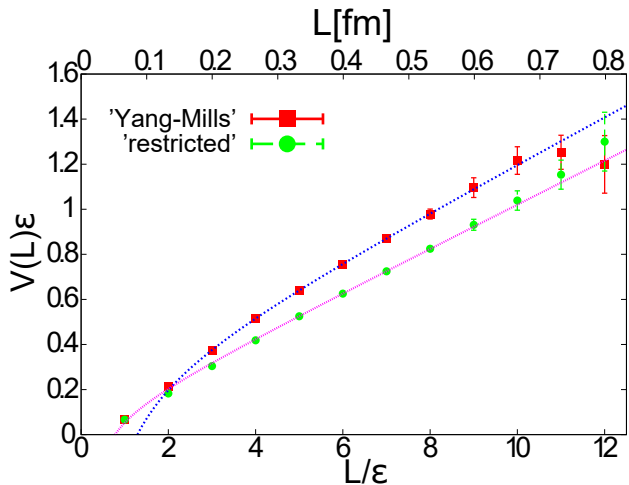}
\caption{
The static potential $V(L,T=6)$ between the sources in the adjoint representation of $SU(2)$ using \cref{su2} for $J=1$ and for comparison the full Wilson loop average in the adjoint representation. The result is consistent with that of \cite{Poulis96,CHS04} where the same quantity is calculated by the Abelian projection method.
The curves are obtained by fitting the data with the Cornel potential.
The fit range is $1\leq L/\varepsilon \leq 8$.
}
\label{fig:su2}
\end{figure}

\begin{figure*}[t]
\centering
\includegraphics[width=0.95\hsize]{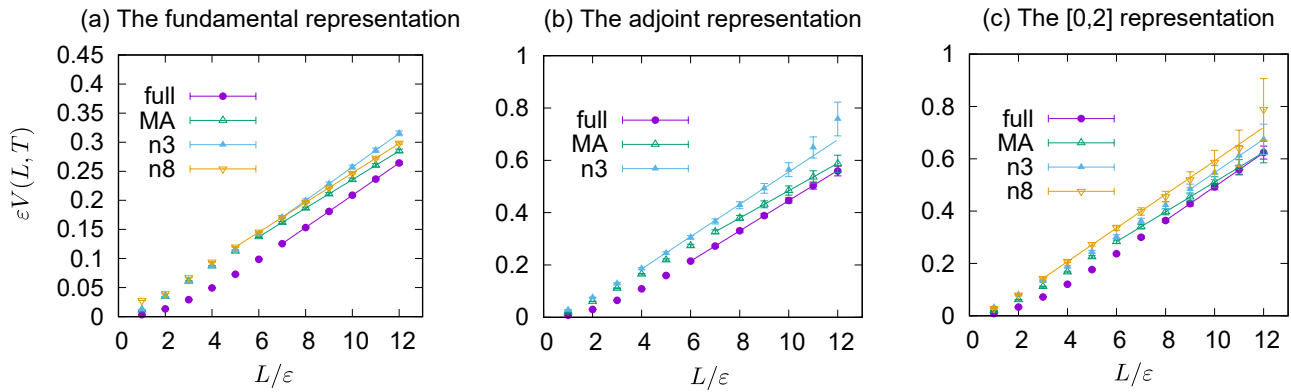}
\caption{
The static potential $\braket{V(L,T=8)}$ between the sources in (a) the fundamental $[0,1]$, (b) adjoint $[1,1]$, and  (c) sextet $[0,2]$ representations of $SU(3)$ calculated using \cref{su3}, in comparison with the full Wilson loop average. The legends, MA, n3 and n8 represents the measurements by using the corresponding reduction conditions Eqs.\ (\ref{MA}), (\ref{n3}) and (\ref{n8}) respectively.
The straight lines are obtained by fitting the data with the linear potential. 
The fit range is indicated by the plotting range of the lines.
}
\label{fig:su3}
\end{figure*}

In order to support our claim that the dominant part of the Wilson loops in higher representation is given by the highest weight part \cref{tildeW},
we examine numerically whether the string tension extracted from \cref{su2,su3} reproduce the full string tension or not.
In this paper we investigate the Wilson loop in the adjoint representation of $SU(2)$ and in the adjoint representation $[1,1]$ and the sextet representation $[0,2]$ of $SU(3)$.

We set up the gauge configurations for the standard Wilson action
at $\beta = 2.5$ on the $24^4$ lattice for $SU(2)$ and at $\beta=  6.2$  on the $24^4$ lattice for $SU(3)$.
For $SU(2)$ case, we prepare 500 configurations   every 100 sweeps after 3000 thermalization by using the heatbath method.
For $SU(3)$ case, we prepare 1500 configurations  evey  50  sweeps after 1000 thermalization   by using pseudo heatbath method with over-relaxation algorithm (20 steps per sweep).
In the measurement of the Wilson loop average we apply the hyper-blocking for $SU(2)$ case and the APE smearing technique for $SU(3)$ case to reduce noises and the exciting modes.
In $SU(3)$ case, the number of the smearing is determined so that the ground state overlap is enhanced \cite{BSS95}.
We have calculated the Wilson loop average $W(L,T)$ for a rectangular loop with length $T$ and width $L$ to derive the potential $V(L,T)$ through the formula
\begin{align}
  V(L,T) = -\log\frac{W(L,T+1)}{W(L,T)}.
\end{align}


In case of $SU(2)$, we investigate the Wilson loop in the adjoint representation $\bm 3$ ($J=1$).
The restricted link variable $V_{x,\mu}$ is obtained by using \cref{lat_decom} for the color field $\bm n_x$ which minimizes the reduction functional \cref{MA} ($N=2$).
\Cref{fig:su2} shows that the static potentials from the proposed operator \cref{su2} for $J=1$ and the full Wilson loop in the adjoint representation are in good agreement.
The string tensions $\sigma_{\mathrm{full}}$ and $\sigma_{\mathrm{rest}}$ for the full Wilson loop and the proposed operator which are extracted by fitting the data with the Cornel potential are
\begin{align}
  &\sigma_{\mathrm{full}} = 0.1021(234), \qquad \sigma_{\mathrm{rest}} = 0.0968(159), \notag\\
  &\sigma_{\mathrm{rest}}/\sigma_{\mathrm{full}} \simeq 0.95.
\end{align}
Note that in the fundamental representation $\bm 2$ ($J=1/2$), we obtain the perfect Abelian dominance in the string tension in \cite{KKS15}.

\begin{table}[t]
\centering
\caption{
The string tensions in the lattice unit in the $SU(3)$ case:  the string tensions obtained under reduction conditions  MA \cref{MA}, n3 \cref{n3} and n8 \cref{n8}, in comparison with the full string tension.
The second line of each cell indicates the ratio of the string tensions which are extracted from the proposed operator and the full Wilson loop for each reduction condition.
Note that the data in the slot $[1,1]$-n8  is not available, because the highest weight part of the Abelian Wilson loop in the adjoint representation $[1,1]$ is not determined by the reduction condition n8 \cref{n8}.
}
\begin{tabular}{c||c|c|c|c} \hline
&full & MA & n3 & n8 \\\hline\hline
$[0,1]$ & $0.02776(2)$ & $0.02458(1)$ & $0.02884(3)$ & $0.02544(3)$ \\
 &&$89\%$ &$104\%$&$91\%$ \\\hline
$[1,1]$ & $0.0576(1)$ & $0.0522(1)$ & $0.062(1)$ & -\\
&&$91\%$ &$108\%$& \\\hline
$[0,2]$& $0.0647(1)$ & $0.05691(9)$ & $0.0635(2)$ & $0.0641(6)$ \\
&&$91\%$ &$98\%$&$99\%$ \\\hline 
\end{tabular}
\label{tab}
\end{table}
In case of $SU(3)$, we investigate the Wilson loop in the fundamental representation $[0,1] = \bm 3$, the adjoint representation $[1,1]=\bm 8$ and the sextet representation $[0,2]=\bm 6$.
For each representation, we measure the Wilson loop average for possible reduction functionals, \cref{MA,n3,n8}.
\Cref{fig:su3} shows the static potentials from the proposed operator \cref{su3} for $[m,n]=[0,1], [1,1], [0,2]$ and the full Wilson loop in the fundamental, adjoint and sextet representations.
\Cref{tab} shows the string tensions which are extracted by fitting the data with the linear potential. 
Note that the data for the adjoint representation $[1,1]$ under the reduction condition n8   is not available, since the highest weight part of the Abelian Wilson loop in the adjoint representation $[1,1]$ is not determined by the reduction condition n8 \cref{n8}, as explained in the final part of the previous section. 
The string tensions extracted from the proposed operator reproduce nearly equal to or more than $90\%$ of the full string tension for any of the reduction conditions \cref{MA,n3,n8}.
These results indicate that the proposed operators give actually the dominant part of the Wilson loop average.

\section{Conclusion} \label{conc}

In this paper, we have proposed a solution for the problem that the correct string tension extracted from the Wilson loop in higher representations cannot be reproduced if the restricted part of the Wilson loop is naively extracted by adopting the Abelian projection or the field decomposition  in the same way as in the fundamental representation.
We have given a prescription to construct the gauge invariant operator \cref{tildeW} suitable for this purpose. 
We have performed numerical simulations to show that this prescription works well in the adjoint representation $\bf{3}$ for $SU(2)$ color group, and the adjoint representation $[1,1]=\bf{8}$ and the sextet representation $[0,2]=\bf{6}$ for $SU(3)$ color group.
In comparison, we have investigated the Wilson loop for the restricted field in the fundamental representation of $SU(3)$, by using the reduction conditions \cref{MA,n3,n8}.
It should be compared to the result of \cite{KSSK11} calculated by using the minimal option, which is a different option of the field decomposition where $V_{x,\mu}$ and $X_{x,\mu}$ are determined by using only $\bm n^8_x$.

Further studies are needed in order to establish the magnetic monopole dominance in the Wilson loop average for higher representations, supplementary to the fundamental representation for which the magnetic monopole dominance was established.
In addition we should investigate on a lattice with a larger physical spatial size because it was stated in \cite{SS14} that for the sufficiently large spatial size, the Abelian part of the string tension perfectly reproduced the full string tension in the fundamental representation of $SU(3)$.
It should be also checked whether the string breaking occurs for the highest weight part of the Abelian Wilson loop in the adjoint representation of $SU(3)$, similarly to $SU(2)$ case \cite{CHS04}.

\begin{acknowledgments}
This work was supported by Grant-in-Aid for Scientific Research, JSPS KAKENHI Grant Number (C) No.15K05042.
R. M. was supported by Grant-in-Aid for JSPS Research Fellow Grant Number 17J04780. 
 The numerical calculations were in part supported by the Large Scale Simulation Program No.16/17-20(2016-2017) of High Energy Accelerator Research Organization (KEK), and were performed in part using COMA(PACS-IX) at the CCS, University of Tsukuba.
\end{acknowledgments}

\appendix
\section{The derivation of \cref{WV}}\label{sec:deriv_WV}
The following derivation can be applied to an arbitrary compact gauge group.
The two conditions which determines the decomposition, \cref{cond1,cond2}, are common to all compact gauge groups.

By gauge-transforming $\mathscr V_\mu$ by $\Theta$ in \cref{cond1} and using \cref{color_lat}, we obtain
\begin{align}
  [\mathscr V_\mu^{\Theta},H_k] = 0,\qquad k=1,\ldots,r,
\end{align}
where $\mathscr V_\mu^{\Theta} := \Theta^\dag V_\mu \Theta + i{g_{\mathrm{YM}}^{-1}}\Theta^\dag\partial_\mu\Theta$.
This means that $\mathscr V_\mu^\Theta$ belongs to the Cartan subalgebra and thus it is commutable with itself, $[\mathscr V_\mu^\Theta(x),\mathscr V_\nu^\Theta(y)] = 0$.
Therefore by transforming $\mathscr V_\mu$ by $\Theta$ in \cref{WV}, we obtain
\begin{align}
  &\frac1{D_R}\tr_R\mathcal P\exp\left(i{g_{\mathrm{YM}}}\oint \mathscr V \right) \notag\\
  &= \frac1{D_R}\tr_R \exp\left(i{g_{\mathrm{YM}}}\oint\mathscr V^\Theta \right), \label{a2}
\end{align}
where we can omit the path ordering because $\mathscr V_\mu^\Theta$ is commutable.
  The trace of an element $\exp(i\phi_k H_k)$ of the Cartan subgroup in $R$ is calculated as
  \begin{align}
    \tr_R\exp(\phi_k H_k) &= \sum_{\bm\mu\in\Delta_R}d_\mu\bra{\mu}\exp(i\phi_k H_k) \ket\mu \notag\\
    &= \sum_{\bm\mu\in\Delta_R}d_\mu\exp(i\phi_k \mu_k) \notag\\
    &= \sum_{\bm\mu\in\Delta_R}d_\mu\exp(i\bra\mu\phi_k H_k \ket\mu),
  \end{align}
  where we have used $H_k\ket\mu = \mu_k\ket\mu$.
Therefore, by performing the trace in \cref{a2}, we obtain
\begin{align}
  (\ref{a2})=\frac1{D_R}\sum_{\bm \mu\in\Delta_R}d_\mu\exp\left(i{g_{\mathrm{YM}}}\oint\bra{\mu}\mathscr V^\Theta\ket{\mu} \right). \label{a3}
\end{align}
Now \cref{cond2} implies
\begin{align}
  &\tr(\mathscr V_\mu^{\Theta}H_k) = \tr(\mathscr A_\mu^{\Theta}H_k) \notag\\
  \Rightarrow &\mathscr V_\mu^\Theta = \kappa\tr(\mathscr A_\mu^{\Theta}H_k)H_k,
\end{align}
where $\kappa$ is the normalization of the trace and the second line follows from the fact that  $\mathscr V_\mu^\Theta$ belongs to the Cartan subalgebra.
Because for an element $E$ of the Lie algebra,
\begin{align}
  \bra{\mu}\tr(EH_k)H_k\ket{\mu} = \bra{\mu}E\ket{\mu},
\end{align}
we obtain
\begin{align}
  (\ref{a3}) =\frac1{D_R}\sum_{\bm \mu\in\Delta_R}d_\mu\exp\left(i{g_{\mathrm{YM}}}\oint\bra{\mu}\mathscr A^\Theta\ket\mu \right).
\end{align}
This completes the derivation of \cref{WV}.

\section{The derivation of \cref{su2,su3}}\label{sec:deriv_tildeW}
First we show \cref{su2} in $SU(2)$ Yang-Mills theory.
The Wilson loop for the restricted field can be written by using Abelian link variables $u_{x,\mu}$ which are defined by
\begin{align}
  u_{x,\mu} := \Theta_x^\dag V_{x,\mu} \Theta_{x+\mu}. \label{Abel_link}
\end{align}
Here it should be noted that an Abelian link variable $u_{x,\mu}$ belongs to the Cartan subgroup $U(1)^{N-1}$ because of \cref{cond1lat}.
The (normalized) trace of the product of the Abelian link variables along a closed loop $C$ is equal to the Wilson loop for the restricted link variables $V_{x,\mu}$ as
\begin{align}
  W_R[V;C] &:= \frac1{D_R}\mathrm{tr}_R \prod_{\braket{x,\mu}\in C}V_{x,\mu}  \notag\\
  &=\frac1{D_R}\tr_R \prod_{\braket{x,\mu}\in C}u_{x,\mu},
\end{align}
where $D_R$ is the dimension of a representation $R$ and $\tr_R$ denotes the trace in $R$.
Now we define the untraced Abelian Wilson loop $w_C$, which belongs to $U(1)$, by using \cref{Abel_link} as
\begin{align}
  w_C := \prod_{l\in C}u_l = \Theta_x^\dag V_C \Theta_x,
\end{align}
where $x$ is the starting point of $C$.
Let us parameterize the untraced Abelian Wilson loop as
\begin{align}
  \Theta_x^\dag V_C \Theta_x = \operatorname{diag}(e^{i\theta},e^{-i\theta}).
\end{align}
Then the proposed operator \cref{tildeW} in the spin-$J$ representation is written as
\begin{align}
  \widetilde W_J[V;C] = \frac12 ( e^{i2J\theta} + e^{-i2J\theta}).
\end{align}
Therefore we obtain
\begin{align}
  \frac12\tr \left((V_C)^{2J}\right) = \frac12 \tr (\Theta_x^\dag V_C \Theta_x)^{2J}
  = \widetilde W_J[V;C]
\end{align}
This completes the derivation of \cref{su2}.

In the $SU(3)$ Yang-Mills theory, let us parameterize the untraced Abelian Wilson loop as
\begin{align}
  w_C := \Theta_x^\dag V_C \Theta_x = \operatorname{diag}(e^{i\theta_1},e^{i\theta_2},e^{i\theta_3}),
\end{align}
where $\theta_1+\theta_2+\theta_3 = 0 \bmod 2\pi$.
Then the proposed operator \cref{tildeW} in the $[m,n]$ representation is written as
\begin{align}
  \widetilde W_{[m,n]}[V;C] = &\frac16 (e^{i(m\theta_1 - n\theta_3)} + e^{i(m\theta_3 - n\theta_1)} \notag\\
 &+e^{i(m\theta_3 - n\theta_2)} + e^{i(m\theta_2 - n\theta_3)} \notag\\
 &+e^{i(m\theta_2 - n\theta_1)} + e^{i(m\theta_1 - n\theta_2)}).
\end{align}
Therefore we obtain
\begin{align}
  &\tr \left((V_C)^{m}\right)\tr \left((V_C^{\dag})^{ n}\right) = \tr \left((w_C)^{m}\right) \tr \left((w_C^{\dag})^{n}\right)\notag\\
  &= (e^{im\theta_1} + e^{im\theta_2} + e^{im\theta_3}) \notag\\
  &\quad\times(e^{-in\theta_1} + e^{-in\theta_2} + e^{-in\theta_3}) \notag\\
  &=6\widetilde W_{[m,n]}[V;C] \notag\\
  &\quad+ e^{i(m-n)\theta_1} + e^{i(m-n)\theta_2} + e^{i(m-n)\theta_3} \notag\\
  &=6\widetilde W_{[m,n]}[V;C] + \tr\left((w_C)^{m}(w_C^{\dag})^{ n}\right) \notag\\
  &=6\widetilde W_{[m,n]}[V;C] + \tr\left((V_C)^{m}(V_C^{\dag})^{ n}\right).
\end{align}
This completes the derivation of \cref{su3}.

\end{document}